\documentclass[traditabstract]{aa}
\usepackage{graphicx}
\usepackage{color}
\usepackage{longtable}
\usepackage{multirow}
\usepackage{natbib}
\usepackage{txfonts}
\usepackage{fixltx2e}
\usepackage{wasysym}
\usepackage{marvosym}

\begin{document}

\def\be{\begin{equation}}
\def\ee{\end{equation}}
\def\vel{$\gamma^2$~Velorum}
\def\g2vel{\gamma^2Vel}
\def\al{$^{26}$Al}

\title{Modeling of the Vela complex including the Vela supernova remnant, the binary system \vel, and the Gum nebula}

\author{ I. Sushch \inst{1,2} \and B. Hnatyk \inst{3} \and A. Neronov \inst{4}}

\institute{
  Humboldt Universit\"{a}t zu Berlin, Institut f\"{u}r Physik, Berlin, Germany
  \and
  National Taras Shevchenko University of Kyiv, Department of Physics, Kyiv, Ukraine
  \and
  National Taras Shevchenko University of Kyiv, Astronomical Observatory, Kyiv, Ukraine
  \and 
  ISDC, Versoix, Switzerland
  }

\date{Received 07 July 2010; accepted 18 October 2010}

\abstract
{We study the geometry and dynamics of the Vela complex including the Vela supernova remnant (SNR),
the binary system \vel\ and the Gum nebula. We show that the Vela SNR belongs to a subclass of
non-Sedov adiabatic remnants in a cloudy interstellar medium (ISM), the dynamics of which 
is determined by the heating and evaporation of ISM clouds. We explain observable
characteristics of the Vela SNR with a SN explosion with energy
$1.4\times 10^{50} \mbox{ ergs}$ near the step-like boundary of the ISM with low intercloud
densities ($\sim 10^{-3}\mbox{ cm}^{-3}$) and with a volume-averaged density of clouds evaporated by
shock in the north-east (NE) part about four times higher than the one in the south-west (SW)
part. The observed asymmetry between the NE and SW parts of the
Vela SNR could be explained by the presence of a stellar wind bubble (SWB) blown by the nearest-to-the 
Earth Wolf-Rayet (WR) star in the \vel\ system. We show that the size and kinematics of \vel\ SWB agree
with predictions of numerical calculations for the evolution of the SWB of $M_{\rm ini} = 35M_\odot$ star. The low initial mass of the
WR star in \vel\ implies that  the luminosity of the nuclear line of $^{26}$Al, produced by \vel, is below
the sensitivity of existing gamma-ray telescopes.}


\authorrunning{I. Sushch et al.}
\titlerunning{Modeling of the Vela complex}
\maketitle

\section{Introduction}

The Vela complex is one of the most interesting regions in the galactic plane. The observable flux
from the Vela region ranges from radio to TeV energies. It consists of many objects, including
the Gum nebula, the Vela supernova remnant (SNR) and Vela Jr. SNR (SNR RX J0852.0-4622 superposed on Vela), the binary system \vel, the IRAS Vela Shell, an OB-association, $\zeta$ Pup,
etc. Some of them are shown in Fig. \ref{fig:Vela_IVS_Gum}. Because of the distances to these objects
it is possible that some of them can intersect not only in projection, but also physically.  We propose a scenario of joint evolutionary interaction of the Vela SNR, the binary system \vel, the IRAS Vela Shell, the Vela OB2-association, and the Gum nebula.

\section{A hydrodynamical model of the Vela SNR evolution}

\subsection{Vela SNR properties and peculiarities. NE-SW asymmetry.}

The Vela SNR is one of the closest supernova remnants to us. Different
estimates of the distance to the Vela SNR suffer from a large
uncertainty: from $250\pm 30$~pc \citep{cha99} to $350$~pc
(\citet{dubner98} and references therein). Hubble Space telescope
parallax observations of the Vela pulsar give the distance to the pulsar of
$D_{\rm Vela}=294^{+76}_{-50}$~pc \citep{caraveo01}, and the best estimate
is from the VLBI parallax measure \citep{dodetal03}:
\begin{equation}
D_{\rm Vela}=287^{+19}_{-17} \mbox{ pc}.
\label{D_Vela}
\end{equation}

Similarly uncertain are the estimates of the Vela
SNR age, which range from a few thousand years \citep{stothers80} to
$t_{\rm SNR}\simeq 2.9\times 10^4$~yr \citep{aschenbach95}. The most
commonly cited estimate is $t_{\rm SNR}=t_{\rm pulsar}\simeq
1.14\times 10^4$~yr, where $t_{\rm pulsar}$ is the age of the Vela
pulsar \citep{reichley70}.
The total $0.1-2.4$ keV X-ray luminosity from Vela SNR in erg/s is \citep{lu00}
\begin{equation}
L_{\rm x, tot}=3.0\times10^{35}\left[\frac{D_{\rm vela}}{290\rm pc}\right]^{2}\mbox{ erg/s}.
\label{Lxtot}
\end{equation}

\begin{figure}

\includegraphics[width=\linewidth]{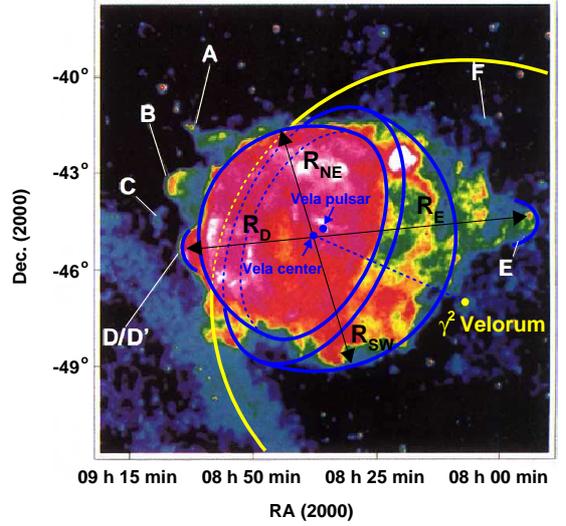}
\caption{ ROSAT All-Sky Survey image (0.1--2.4 KeV) of the Vela SNR \citep{aschenbach95}. A--F are extended features outside the boundary of the remnant ("bullets"). Light blue to white contrast represents a contrast in surface brightness of a factor of 500 \citep{aschenbach95}. Blue curves show the NE and SW hemispheres of the Vela SNR. The yellow curve shows the contour of the SWB of \vel.} 
\label{vela}
\end{figure}

The main peculiarity of Vela SNR is the
difference in the X-ray brightness and radius of its south-west (SW) and north-east (NE) parts.
The {\it ROSAT} All-Sky Survey image (Fig. \ref{vela}) of the Vela SNR reveals a shell
with a diameter of about $8^\circ$ \citep{aschenbach95}, which implies
a mean linear diameter of
\begin{equation}
d_{\rm Vela}\simeq 40\left[\frac{D_{\rm Vela}}{290\mbox{
pc}}\right]\mbox{ pc}.
\end{equation}
The SW part of the shell
appears to have a radius larger by a factor of
\begin{equation}
R_{\rm SW}\simeq 1.3 R_{\rm NE}
\end{equation}
than the NE part.

Apart from the
difference of the radii, the spatially-resolved spectroscopic
analysis by \citet{lu00} shows that the SW part of the shell appears
to be hotter than the NE one. The shell is
bright only on the NE side, while the SW 
side appears to be dim and is apparently more extended (see Fig.
\ref{vela}). The boundary between the bright and the dim part of
shell is quite sharp \citep{lu00}. The change in the properties of
the shell at different sides indicates that the characteristics of
the ISM in the NE part differ from those in the SW
part. The sharpness of the boundary shows that the change in the ISM properties is abrupt rather than gradual.

\citet{lu00} estimate the contrast change of the two faint regions, each $1.5^\circ \times 1.5^\circ$, in the NE and SW parts, to be a factor of $\simeq 11$ in brightness and $\sim 6$ in emission measure.  Assuming the same estimate for the emission measure contrast change for the entire NE and SW parts of the SNR and taking into account the difference of the radii of the two parts of the SNR, one can find that the luminosity of the NE part $L_{x,NE}$ is a factor of  $\sim 3.7$ higher than the luminosity $L_{x,SW}$ of the SW part.

Apart from the difference in the overall luminosity and temperature, the characteristics of the X-ray emission from the NE and SW parts of the SNR also reveal a difference in the column density of the absorbing material along the line of sight \citep{lu00}. The absorption column densities $N_{\rm H}$ range  from $5.0\times10^{19}$ cm$^{-2}$ in the NE part to $6.0\times10^{20}$ cm$^{-2}$  in the SW part.

Finally, the Vela SNR is peculiar in still another aspect: the main shock of the SNR is not observed. Instead, the bulk of the X-ray emission is distributed all over the SNR volume. Such an observational appearance can be caused by the SNR expanding into a highly inhomogeneous ("cloudy") medium. In this case, the main shock advances through a low-density interstellar medium (ISM), leaving behind denser clouds, which are subsequently heated and partially evaporated by thermal conductivity and transmitted shocks. This results in the appearance of a distributed emission throughout the entire volume of the SNR, instead of from a thin shell at the interface of the main shock with the low-density ISM. The main X-ray emitters in the remnant are the two (cool and hot) phases (components) of heated cloud matter \citep{lu00, miceli05,miceli06}.

\subsection{A hydrodynamical model of the Vela SNR }

Despite extensive investigations, there is no common agreement about the evolutionary status of the Vela SNR. In early studies, a distance of 500 pc was assumed, which corresponds to a rather large size of 70 pc in diameter. Together with this assumption, the absence of the X-ray limb-brightening effect and of the clear signature of the high-speed shock in the SNR
suggested  a radiative stage of the Vela SNR evolution. The appearance of filament structures 
in optic and radio waves as well as 100 km/s absorption lines in the spectra of background 
stars additionally supported this scenario (see, e.g. \cite{gvaramadze99} and references 
therein). But recent observational data strongly support the hypothesis of the Vela SNR 
being in the adiabatic stage. New results for the distance place the Vela SNR  at $D\approx290$ pc, 
implying that the mean radius of the Vela SNR is $R\approx20$ pc, and the dynamics of bullets (ejecta 
fragments) outside the SNR boundary suggest the expansion of the SNR in the low-density ISM. The 
presence of the Vela pulsar indicates that the Vela SN progenitor was of $M_{ini}\leq 25$ 
$M_{\odot}$. For $M_{ini}=11 - 25$ $M_{\odot}$ presupernova masses are $M_{fin} =10.6 - 16.6$ 
$M_{\odot}$ and masses of the ejecta $M_{ej}=9 \div 15$ $M_{\odot}$. (\cite{limongi06}, 
\cite{eldridge04}, \cite{kasen09}). Therefore, the  interaction (merger) of the massive 
ejecta with a velocity of over 1000 km/s with the RSG shell occurs in the adiabatic regime, 
and the Vela SNR with a radius of 20 pc and an age of about 10000 years should be in the 
adiabatic stage of evolution with characteristic velocities of about $mR_{SNR}/t_{SNR}\approx$ 
1000 km/s ($m\leq 1$ in the free expansion case and $m=0,4$ in the adiabatic one) without 
forming a thin dense radiative shell. Another observational confirmation of the absence of 
the 100 km/s shell follows from the studies of absorption lines in the spectra of background 
stars \citep{cha00}, where all stars in the Vela SNR direction at distances smaller than 350 
pc do not show evidences of a 100 km/s absorption line. The maximum broadening corresponds  
to $<50$ km/s. Only stars with distances exceeding 500 pc show 100 km/s features. Radio and 
optical shells (shell-like filaments) can be naturally explained by the emission of 
filamentary structures, exited by the SNR shock (akin to the Cyg Loop case, where 
filaments coexist with the main fast adiabatic shock).

The hydrodynamical model of the Vela SNR evolution cannot be described directly by the Sedov solution \citep{sedov59},
because the expansion proceeds in a cloudy rather than homogeneous ISM. Instead, to describe the Vela SNR evolution one can use the White \& Long solution, which describes the evolution of a supernova remnant
expanding into a cloudy ISM \citep{white}.

As explained above, the observed asymmetry of the Vela SNR is most probably due to the difference in
the properties of the ISM on different sides of the remnant, in particular, by the different densities of clouds
or different density contrast between the clouds and the IC medium. Below we assume that before the SN explosion the average
number density (concentration) and, therefore, volume filling factor of clouds in NE part was larger than that
in the SW part. This naturally explains the smaller radius and the lower temperature of the NE part of the remnant.  Once more, interaction of shock wave with clouds results in numerous filamentary structures of disrupted cloud material, visible in radio band, and the radio image of the Vela SNR really shows more numerous shell structures in the NE part than in SW one \citep{bock98}. The NE-SW asymmetry proposed here can also explain a pulsar wind nebula (PWN) displacement to the SW according to the pulsar position. Namely, the reverse shock from more dense NE part reaches and destroys PWN earlier \citep{blondin01, lamassa08}       

The conjecture that the densities on the two sides of the SNR are different is supported also by the
dynamics of the shrapnel (bullets) - high-velocity clumps of the SN ejecta with an overabundance of heavy elements \citep{aschenbach95, miceli08, lamassa08, yamaguchi09}. X-ray spectra of shrapnel argue a low density of the ISM around the Vela SNR, whereas a difference in the distances traveled by the  protruding shrapnel on the NE (shrapnel pieces A-D/D') and on SW (shrapnel pieces E and F) sides of the SNR, as it is
illustrated for the shrapnel pieces D and E shown in Fig. \ref{vela}, is consistent with a corresponding difference in mean ambient  density, assuming that each shrapnel piece has approximately the same density and initial velocity.

The  solution for the shock radius $r_{s}$ of the remnant as a function of the age $t$ has the
form \citep{white}
\begin{equation}
\label{rs}
r_{s}=\left[\frac{25(\gamma+1)KE}{16\pi\rho_{\rm ic}}\right]^{1/5}t^{2/5},
\end{equation}
where $E$ is the explosion energy, $\rho_{\rm ic}$ is the density of the  intercloud medium and $\gamma=5/3$ is the adiabatic index. The solution (\ref{rs}) differs from the standard Sedov solution without
evaporating clouds by a different choice of a phenomenological constant $K$.

The constant $K$ depends on two parameters:
on the ratio $C$ of the fraction of the ISM mass initially contained in clouds and evaporated behind the shock wave to the ISM mass in the intercloud medium
\begin{equation}
C=\frac{\left<\rho_{\rm c}\right>}{\rho_{\rm ic}},
\end{equation}
where $\left<\rho_{\rm c}\right>$ is the volume-averaged density of the clouds, and on the ratio of the cloud evaporation time
scale $t_{\rm ev}$ to the SNR age $t$
\begin {equation}
\tau={t_{\rm ev}}/{t}.
\end {equation}
To express the dependence of $K$ on these two parameters we can use a simple analytical approximation
\begin {equation}
\frac{K}{K_S}\simeq\left(1+\frac{C}{1+\tau}\right)^{-1},
\end {equation}
and $K/K_S\sim C^{-1}$ for $C\ll1,\tau\gg1$, where $K_S=1.528$ is the value of constant $K$ for the Sedov solution (when $C=0$). 
From eq. (\ref{rs}) follows the value of the shock velocity
\begin {equation}
V_{\mbox{s}}=\frac{2}{5}\frac{r_s}{t}=\left[\frac{(\gamma+1)K E}{4\pi\rho_{\rm ic}r_{s}^{3}}\right]^{1/2},
\label{Vs}
\end {equation}
which allows to calculate the temperature at the shock
\begin {equation}
T_{S}=\frac{2(\gamma-1)}{(\gamma+1)^2}\frac{\mu m_{H}}{k}V_{\mbox{s}}^{2}
\label{Ts}
\end {equation}
(here $\mu=16/27$ is the average mass per particle in hydrogen mass units $m_H$  for helium/hydrogen abundance ratio $n_{\rm He}=(1/12)n_{\rm H}$ and $k$ is the Boltzmann constant) for the \citet{white} solution.

The dynamics of expansion of the SNR is fully characterized by a set of four parameters: the explosion energy $E$, the preshock intercloud ISM density $\rho_{\rm ic}$, the cloud/intercloud density ratio $C$ and the evaporation time/SNR age ratio $\tau$. The values of these parameters can be derived from the set of the observed  characteristics of Vela SNR, such as the shock radii, characteristics of the X-ray radiation for the both NE and SW parts, etc.

 For the NE and SW shock radii $R_{\rm NE}\simeq 18$~pc and $R_{\rm SW}\simeq 23$~pc and SNR age
 $t_{\rm Vela}\simeq 1.14\times 10^4$~yr from eq. (\ref{Vs}) we obtain corresponding shock velocities
\be
V_{\rm NE}=0.4R_{\rm NE}/t_{\rm Vela}\approx 6.0\times 10^7\mbox{ cm/s},
\ee
\be
V_{\rm SW}=0.4R_{\rm SW}/t_{\rm Vela}\approx 7.7\times 10^7\mbox{ cm/s},
\ee
and from eq.(\ref{Ts}) shock temperatures $T_S^{NE}\approx 4.8\times 10^6\mbox{K}$,
$T_S^{SW}\approx 7.8\times 10^6\mbox{K}$. Average (emission measure weighted) temperature of plasma
inside the SNR is about twice as high (\citet{white}):
\be
<T^{NE}_{hot}>\approx 1.9T_S^{NE}\approx 9\times 10^6\mbox{K},
\ee
\be
<T^{SW}_{hot}>\approx 1.9T_S^{SW}\approx 1.5\times 10^7\mbox{K}.
\ee

According to the results of \cite{lu00} X-ray radiating plasma in both parts of the Vela SNR
consists of two phases: a hot one ($T \sim 0.5-1.2$ keV) and a cool one ($T \sim 0.09-0.25$ keV),
and the cool one dominates the X-ray luminosity of the SNR. The temperatures found above correspond to
the hot component in the two-temperature Raymond-Smith thermal plasma model used in \cite{lu00}).
The hot evaporated gas component with the volume filling factor $f_{hot}\simeq 1$
dominates the shock dynamics, while the cool one with $f_{cool}=1-f_{hot}\ll1$ dominates in X-ray
radiation. The role of the initial intercloud interstellar gas is negligible in both shock
dynamics and X-ray radiation. It means that
\be
\frac{C_{hot}}{1+\tau}\gg1,
\label{Chot}
\ee
where $C_{hot}=<\rho_{c,hot}>/\rho_{ic}$, and hereafter we take $\tau \ll 1$ and use the approximation
$K/K_S=C_{hot}^{-1}$. In this case, eq.(\ref{rs}) for shock radius is reduced to
\be
r_{\rm Vela}=
              \left[\frac{25(\gamma+1)K_SE}
             {16\pi m_H n_{hot}}\right]^{1/5}t^{2/5},
\label{rv}
\ee
where $n_{hot}=\rho_{c,hot}/m_H$ is the nucleon number density of the hot component. As we can see,
Eq.(\ref{rv}) is the Sedov solution, in which a mean density of intercloud plasma inside the remnant
$\rho_{\rm ic}$ is replaced by the mean density of the evaporated clouds (of the dominant hot component
$\rho_{c,hot}$ in our case).

The nucleon number density of the hot component $n_{hot}$ can be estimated from its X-ray radiation.
The X-ray luminosity of the SNR is an integral over SNR volume $\cal V$ for X-ray emissivity
$\epsilon_X=n_e n_H \Lambda_X(T)$ of plasma with temperature $T$, electron (hydrogen) number density
$n_e(n_H)$ and cooling function $\Lambda_X(T)$
\be
\label{Lx}
L_{X}=\int_{\cal V} \epsilon_X d\cal V .
\ee
Within the model of \citet{white}, $L_{X}$ is estimated in erg/s as
\begin {eqnarray}
\label{Lx_w}
L_{X}=1.7\times10^{34}Q\Lambda_{-22}\left[\frac{n_{\rm ic}}{1\mbox{ cm}^{-3}}\right]^{2}
\left[\frac{r_{s}}{1\mbox{ pc}}\right]^{3}\left[1+\frac{C}{1+\tau}\right]^2,
\end {eqnarray}
where  $n_{\rm ic}=\rho_{\rm ic}/m_H$ is the intercloud ISM nucleon number density (nucleon number density $n=(4/3)n_{\rm H}$  )), $\Lambda_{-22}$
is  the cooling function in units $10^{-22}\,\mbox{erg}\,\mbox{cm}^{3}\,\mbox{s}^{-1}$,
$Q$ is a number on the order of one and depends on $C$ and $\tau$ (for $C\gg1$ and $\tau\ll1$ $Q\simeq 1$).

Unfortunately, we do not know the X-ray luminosity of the hot component of Vela SNR with satisfactory
accuracy. Therefore, we use the more accurate data about the emission measure of the hot plasma for the
calculation of the nucleon number density of the hot component. According to the observation of
\citet{lu00}, the mean value of the emission measure A of
$\varphi_{pixel}\times\varphi_{pixel}=8.75'\times 8.75'$ pixel
\be
A=10^{-14}\int_{pixel} n_e n_H d{\cal V}/4\pi D_{\rm Vela}^2
\label{Ap}
\ee
of the hot plasma in NE region is $A^{NE}_{hot}\simeq (5\div 7)\times 10^{-4} \mbox{ cm}^{-5}$.
Meanwhile, from eq. (\ref{Ap}) and conditions $n_en_H = 0.66n^2$,
${\cal V}_{pixel}=(\varphi_{pixel}D_{\rm Vela})^2\times 2R_{SNR}$ it follows that
\be
 nf^{1/2}=1.5A^{1/2}\left[\frac{R_{\rm Vela}}{20\mbox{ pc}}\right]^{-1/2}\,cm^{-3}
\label{nf}
\ee
and, taking into account that $f_{hot}^{NE}\simeq 1$,
\be
n_{hot}^{NE}\simeq 4.0\times 10^{-2}\mbox{ cm}^{-3}.
\label{nhne}
\ee

Now, from eq.(\ref{rv}), we can find nucleon number density of hot component in SW part of Vela SNR
\be
n_{hot}^{SW}=n_{hot}^{NE}\left[\frac{R_{\rm NE}}{R_{\rm SW}}\right]^5
             \simeq 1\times 10^{-2}\mbox{ cm}^{-3}
\label{nhsw}
\ee
and the energy of the Vela SNR explosion
\be
E=\frac{16\pi m_H }{25(\gamma+1)K_S}\frac{R_{\rm NE}^5}{t_{\rm Vela}^2}n_{hot}^{NE}
\simeq 1.4\times 10^{50}\mbox{ erg}.
\label{E}
\ee

The cool component of the X-ray radiating plasma in the Vela SNR does not influence the shock dynamics, but dominates in the X-ray luminosity. The mean value of the emission measure $A$ and the temperature of the cool plasma in the NE region is $A^{NE}_{cool}\simeq (3\div 5)\times 10^{-3} \mbox{ cm}^{-5}$,
$T_{cool}^{NE}\simeq 0.1\times 10^7 \mbox{ K}$ and in the SW region $A^{SW}_{cool}\simeq (4\div 6)\times 10^{-4} \mbox{ cm}^{-5}$, $T_{cool}^{SW}\simeq 0.17\times 10^7 \mbox{ K}$ \citep{lu00}. From eq.(\ref{nf}) for the NE part of the Vela SNR follows
\be
(n_{cool}f_{cool}^{1/2})^{NE}=1.6(A_{cool}^{NE})^{1/2}\mbox{ cm}^{-3}
                              \simeq 1.0\times 10^{-1} \mbox{ cm}^{-3}.
\label{nfcne}
\ee
The filling factors of hot and cool plasma in the NE part can be estimated from the equality of the pressure $P\propto nT$ of both components: $ (n_{cool}T_{cool})^{NE}\simeq (n_{hot}T_{hot})^{NE}$
\be
\frac{f_{cool}^{NE}}{f_{hot}^{NE}}=\frac{f_{cool}^{NE}}{1-f_{cool}^{NE}}=
\left[\frac{T_{cool}^{NE}(n_{cool}f_{cool}^{1/2})^{NE}}
{T_{hot}^{NE}(n_{hot}f_{hot}^{1/2})^{NE}}\right]^2\simeq 8\times 10^{-2}
\label{fcne}
\ee
or $f_{cool}^{NE}\simeq 0.07$, $f_{hot}^{NE}\simeq 0.93$. And, finally, from eq.(\ref{nf}) follow nucleon number densities  $n_{cool}^{NE}\simeq 0.38 \mbox{ cm}^{-3}$, $n_{hot}^{NE}\simeq 0.04 \mbox{ cm}^{-3}$.

Similarly, for the SW part of the Vela SNR $(n_{cool}f_{cool}^{1/2})^{SW}\simeq 3.1\times10^{-2}\mbox{ cm}^{-3}$, $f_{cool}^{SW}\simeq 0.09$, $f_{hot}^{SW}\simeq 0.91$, and $n_{cool}^{SW}\simeq 0.10 \mbox{ cm}^{-3}$, $n_{hot}^{NE}\simeq 0.01 \mbox{ cm}^{-3}$.

Table 1 gives a summary of the parameters of the NE and SW parts of the SNR derived from the X-ray data.

As a test for the self-consistency of our model we can calculate the predicted X-ray luminosities of
different parts/components of Vela SNR in erg/s, using eq.(\ref{Lx}):
\be
L_{cool}^{NE}=\frac{2}{3}\pi R_{NE}^3f_{cool}^{NE}(n_en_H \Lambda_X(T))_{cool}^{NE}
           \simeq 2.2\times 10^{35}\Lambda_{-22},
\ee
where we use the approximation validated by  \cite{white} $\Lambda_{-22}(T)=1$.
Similarly,
\be
L_{hot}^{NE}\simeq 0.4\times 10^{35}\Lambda_{-22}\mbox{ ergs}^{-1},
\ee
\be
L_{cool}^{SW}\simeq 0.6\times 10^{35}\Lambda_{-22}\mbox{ ergs}^{-1},
\ee
\be
L_{hot}^{SW}\simeq 0.06\times 10^{35}\Lambda_{-22}\mbox{ ergs}^{-1}.
\ee
The total model luminosity $L_{X,tot}\simeq 3.2\times 10^{35}\Lambda_{-22}\mbox{ ergs}^{-1}$
is close to the observed one of eq. (\ref{Lxtot})

\begin{table}
\caption{Characteristics of Vela SNR}
\begin{tabular}{l c c}

\hline
\hline
Parameter&   NE& SW\\
\hline
$E$, [erg]& \multicolumn{2}{c}{$0.14\times10^{51}$}\\

$t$, [years]&\multicolumn{2}{c}{$11400$}\\

$R$, [pc]                            &  $18$     & $23$\\

$V_{\mbox{s}}$, [km/s]               & $600$     & $770$\\

$T_S$, [K]                           & $4.8\times10^{6}$ &  $7.8\times10^{6}$\\

$n_{hot}$,              [cm$^{-3}$]  & $0.04$   & $0.01$\\

$f_{hot}$                          & $0.93$    & $0.91$\\

$T_{hot}$, [K]                       & $9\times10^{6}$ &  $1.5\times10^{7}$\\

$n_{cool}$,              [cm$^{-3}$]  & $0.38$   & $0.10$\\

$f_{cool}$                         & $0.07$    & $0.09$\\

$T_{cool}$, [K]                      & $1\times10^{6}$ &  $1.7\times10^{6}$\\

\hline

\end{tabular}
\end{table}

\section{Interaction of the Vela SNR and the $\gamma^2$~Velorum stellar wind bubble }

Estimates of the physical parameters of the Vela SNR derived in the previous section
show that it is evolving in the inhomogeneous medium whose density changes in a step-like
manner. In this section we explore the possibility that this step-like change of the ISM
parameters can be related to the presence of the boundary of the stellar wind bubble (SWB) around
a Wolf-Rayet (WR) star in the \vel\ system, which is situated in the vicinity of the Vela SNR.

\subsection{Physical characteristics of the \vel\ binary system.}

\vel\ is a WC8+O8-8.5III binary system. The WR
component in this binary system is the closest to Earth WR star (WR11).
The estimates of the distance to \vel\ have evolved over the years and are still controversial.
An early estimate of the distance based on HIPPARCOS parallax was
$258^{+41}_{-31}$~pc \citep{schaerer97}. It was recently revised by several scientific groups. \citet{millour07} give an interferometric estimate of the distance $368_{-13}^{+38}$ pc.  \citet{north07} estimate the distance to \vel\ to be $336^{+8}_{-7}\mbox{ pc}$ based on the orbital solution for the \vel\ binary obtained from the interferometric data. Finally, a revision of the analysis of HIPPARCOS data gives a distance of $334^{+40}_{-32}\mbox{ pc}$ \citet{vanleeuwen07}. Therefore, in the following we take
the distance to \vel\ to be $D_{\g2vel}\simeq 330 \mbox{ pc}$.

The current mass estimate of the
WR star is $M_{\rm WR}=9.0\pm 0.6M_{\odot}$ \citep{north07}. The mass
of the O star is $M_{\rm O}=(28.5\pm 1.1)M_{\odot}$ \citep{north07}.
The \vel\ system is an important source from the viewpoint of nuclear
gamma-ray astronomy. Because it is the nearest WR star, this is the only
source that can potentially be detected as a point source of
1.8~MeV gamma-ray line emission from the radioactive $^{26}$Al with
current generation instruments. Previous observations by COMPTEL put
an upper limit at the level of $1.1\times
10^{-5}$~$\gamma$ cm$^{-2}$s$^{-1}$ on the line flux from the source
\citep{oberlack00}, which is, apparently, below the typical
predictions of the models for stars with the initial mass $M_{\rm
ini}\sim 60M_\odot$.  One should note, however, that modeling of \al\
production suffers from uncertainties of the nuclear reaction
cross-sections, stellar parameters (such as rotation and
metallicity), etc. \citep{limongi06,palacios05}.

The estimate of the initial mass of the WR star, $M_{\rm ini}\sim
(57\pm 15)M_\odot$, is obtained from the evolutionary models
of isolated rather than binary stars \citep{schaerer97}. It is
possible that within a binary system, the mass transfer between the
companions can change the stellar structure of both components, so
that, for example, stars with initial masses as low as $20M_\odot$
can become WR stars in binary systems \citep{vanbeveren91}. Modeling
the binary evolution of \vel\ leads to a lower limit on the
initial mass of the WR star: $M_{\rm WR,i}\ge 38M_\odot$ in
\citep{vanbeveren98} and $M_{\rm WR,i}\simeq 35M_\odot$, $M_{\rm O,i}\simeq  31.5M_\odot$
in \citep{eldridge09}.

\subsection{The stellar wind bubble around \vel}

WR stars are expected to be surrounded by the multi-parsec scale bubbles blown by the strong stellar
wind inside a photoionized HII region.  The size of the HII region and bubble depends on the (time-dependent) joint action of
the photon luminosity in the Lyman continuum $L_{LyC}$ and the mass-loss rate of the star, $\dot M_w$, on its age, $t$, on the wind velocity $v_w$ and on the
density of the ambient medium into which the bubble expands,
$\rho_0$. Qualitatively, the relation between the bubble radius
$R_{\rm bub}$ and the above parameters can be found using the
analytical calculation of \citet{weaver77}
\begin{equation}
\label{weaver}
  R_{\rm bub}=\kappa\left(\frac{1}{2}\dot M_w v_w^2\right)^{1/5}\rho_0^{-1/5}t^{3/5},
\label{rbub}
\end{equation}
where $\kappa$ is a numerical coefficient which, in the simple case
considered by \citet{weaver77} was
$\kappa=\left[125/(154\pi)\right]^{1/5}$, while in a realistic case
it can be found from numerical modeling based on a stellar
evolution model. For example, a  numerical model considered by
\citet{arthur06}, which studied the  evolution of a star with initial mass $40M_\odot$ in a
medium of density $n_{0}=15$~cm$^{-3}$, predicts the final radius of the
shock in the ambient interstellar medium  $R_{\rm sh}\simeq 36$~pc at the end of WR phase before SN explosion.
Using the Eq. (\ref{weaver}) one can re-scale the
numerical simulations for the particular values of ISM density to find that for the
typical value $n_{0}=0.1$~cm$^{-3}$ in the ISM the radius of the SWB
can reach  100~pc. Meanwhile, as we will see later, \vel\ system was born in a molecular
cloud and its bubble should be considerably smaller.

\subsection{Overlap between the Vela SNR and \vel\ bubble}

Comparing the distances of Vela SNR ($\sim 290$pc) and \vel\ system ($\sim 330$pc) and taking into account 
the proximity of the two objects on the sky, one can notice that if the SWB around the 
\vel\ system has indeed a radius of $30-70$~pc ($\ge 5^\circ$ on the sky), the Vela SNR, which itself has a 
size of $\sim 40$~pc, is expected to physically intersect with the \vel\ bubble. In view of this 
geometrical argument it is natural to ascribe the observed step-like change in the parameters of the 
ISM at the location of the Vela SNR to the boundary of the SWB of \vel.

The hypothesis of intersection between the Vela SNR and \vel\ SWB is further supported by the simple geometrical form of the boundary between the bright and the dim part of the SNR shell. Indeed, the boundary roughly follows the contour of an ellipse whose major axis is perpendicular to the direction from the center of Vela SNR toward the \vel, so that the minor axis is aligned with the direction toward \vel, see Fig. \ref{vela}.
The projected distance between the Vela SNR and \vel\ is $D^{\,\prime}=5.2^{\circ}$. Assuming the distances $D_{\rm Vela}\simeq 290 \mbox{ pc}$
and $D_{\rm \g2vel}\simeq 330\mbox{ pc}$  one finds that the physical distance between the two objects is
\begin{equation}
\label{vela-velorum}
R_{\g2vel} = D_{\rm Vela - \g2vel}\simeq 44\mbox{ pc.},
\end{equation}
which we adopt as an estimate for the radius of the SWB around \vel.

\subsection{Estimate of the total mass of the stellar wind bubble of \vel}

Within the geometrical model discussed above, the observed difference in the absorption
column density $N_H$ between the NE and  SW parts of the Vela SNR can be used to estimate
the total mass of the ISM
swept up by the stellar wind of \vel\ over the entire lifetime of the SWB.
Taking the difference between the measured $N_H$ values in the NE part and SW parts  \citep{lu00}
\begin{equation}
\Delta N_H=N_{H,\rm SW}-N_{H,\rm NE}\simeq  5.5\times 10^{20}\mbox{ cm}^{-2}
\end{equation}
 one can find the total mass of the \vel\ SWB:
\begin{equation}
\label{mass}
M\simeq 4 \pi R_{\g2vel}^2 \Delta N_H\times \frac{4}{3}m_H\simeq
             1.3\times10^{5}\mbox{ M$_{\odot}$} \left[\frac{R_{\g2vel}}{44\mbox{ pc}}\right]^{2}.
\end{equation}

Assuming that the bubble has expanded into a homogeneous ISM over the entire expansion history, one would estimate the ISM density around \vel\ as

\begin{equation}
\label{density}
n_{\rm ISM}=\frac{M}{(4/3)\pi R_{\g2vel}^3 m_H}\simeq12\left[\frac{R_{\g2vel}}{44\mbox{ pc}}\right]^{-1}\mbox{ cm}^{-3}.
\end{equation}
One could notice that this estimate of the density of the ISM is much higher than the estimates of both the intercloud and of the volume-averaged cloud density around Vela SNR (Table 1). We come back to the discussion of the origin of this discrepancy below.

\section{The $\gamma^2$~Velorum system and the IRAS Vela shell}

The angular size of the \vel\ SWB, found from our interpretation of the asymmetry 
of the Vela SNR, i.e., from the ratio of $R_{\g2vel}\simeq 44\mbox{ pc}$ and the distance 
$330\mbox{ pc}$ as $\simeq 7.6^\circ$ coincides  with the angular size of a large circular 
arc like structure, centered on $(l,b)=(263^{\circ},-7^{\circ})$ with the radius 
$R_{\rm IVS}\simeq 7.5^{\circ}$, which is visible in the infrared band, known as the "IRAS Vela shell" 
(IVS) \citep{sahu92}. This structure surrounds the Vela OB2 association including \vel\ 
and $\zeta$ Puppis. A recent study of the spatial distribution of the neutral hydrogen and 
radio continuum emission at 1420 MHz  of the IVS by \citet{testori06} provides a new 
estimate of the coordinates of the centroid of IVS from the observed IR emission  
$(l,b)=(259.9^{\circ},-8.3^{\circ})$ and the radius of the neutral hydrogen shell 
in the SW sector between position angles $\sim 162^{\circ}$ and $\sim 265^{\circ}$ 
$R_{\rm IVS}\simeq 5.7^{\circ}$. \citet{testori06} also estimate the mass of ionized 
and atomic components of the shell, assuming a distance of 400 pc: $M_{IVS}\simeq 9.1\times10^{4} M_{\odot}$
(or $M_{IVS}\simeq 6.0\times10^{4} M_{\odot}$ for our distance $330\mbox{ pc}$). It is expected 
that the amount of molecular gas in the IVS is about $\sim 10^5M_{\odot}$(\cite{raj98}).

A possible interpretation of the  IVS as a boundary of the SWB of \vel\ was discussed by \citet{oberlack00} and \citet{testori06}.

Adopting this interpretation, one can find that the estimates of the parameters of the SWB of \vel\, derived above from the analysis of X-ray data on the Vela SNR, agree well with those found from the analysis of the IVS data. In particular, adopting an estimate of the distance toward \vel\ for the center of the IVS, one finds the radius
\begin{equation}
\label{IVS}
R_{\rm IVS}\sim 44\left[\frac{D_{\rm Velorum}}{330\mbox{ pc}}\right]\mbox{ pc}
\end{equation}
and the total mass of the shell
\begin{equation}
M_{IVS}\sim 1\times10^{5} \mbox{ M}_{\odot}
\end{equation}
comparable to the estimates found in Eqs. (\ref{vela-velorum}) and (\ref{mass}), respectively.

\subsection{Implications for the models of the Wolf-Rayet star}

The initial mass of the  WR11 star $M_{\rm ini}$ is equal the sum of the current mass of
the star $M_{\rm WR}=9.0\,M_{\odot}$ and the mass which was blown by the wind:
\begin{equation}
M_{\rm ini}=M_{\rm WR}+\Delta M_{\rm MSS}+
               \Delta M_{\rm (RSGS/LBVS)} +\left<\dot{M_{\rm WR}}\right>t_{\rm WR},
\label{Mini_main}
\end{equation}
where $\dot{M_{WR}}$ is the mass loss rate and $t_{WR}$ is the time interval of the WR wind.
The sum in the above equation includes mass loss via different types of winds ejected by
the star at different stages of stellar evolution: the main sequence stage (MSS) wind, the red
supergiant stage (RSGS) (for stars with initial mass $M_{\rm ini}\leq 40 M_{\odot}$) or the
luminous blue variable stage  (LBVS) (for $M_{\rm ini}\geq 40 M_{\odot}$) wind, and the continuing WR wind. Numerical calculations from \cite{freyeretal03}, \cite{freyeretal05}, \cite{vanmarle05},
\cite{arthur06}, \cite{vanmarle07}, \cite{perez09} show that the MSS and RSGS/LBVS dominate the
mass loss with typical values of mass loss before WR stage $(21\div 26)\,M_{\odot}$ for
$M_{\rm ini}\sim (30\div 60)\,M_{\odot}$. Meanwhile, the WR stage dominates in the kinetic energy of the wind,
injected into the wind bubble.

For the \vel\ binary one can calculate the total stellar
wind mass loss $\Delta M\simeq 29 M_\odot$ as the difference between the initial and contemporary masses of stars in binary systems, which are  $M_{\rm WR,ini}\simeq 35M_\odot$, $M_{\rm O,ini}\simeq  31.5M_\odot$ and $M_{\rm WR}\simeq 9.0M_\odot$, $M_{\rm O}\simeq  28.5M_\odot$ respectively, according to \cite{eldridge09}. This includes the mass-loss rate from the slow red supergiant stage wind $\Delta M_{\rm RSGS}\simeq 19M_\odot$ (\cite{freyeretal05}).
Therefore, the remaining $10 M_{\odot}$ of the hot intercloud gas in the \vel\ SWB correspond to fast MSS and WR winds.
Clumps, created by the interaction of the fast WR wind and the slow RSG wind, together with ISM clouds (Gum nebula interior, see below), survived the passage through the expanded IVS, are the main sources of X-ray emitting plasma in SW part of Vela SNR. 
As follows from Table 1, the lower limit on the total mass of gas in clumps and clouds inside the SWB is
\be
M_{cl}^{SWB}=M_{hot}^{SWB}\left[\frac{n_{hot}^{SW}}{n_{ic}^{SWB}}f_{hot}^{SW}+
                  \frac{n_{cool}^{SW}}{n_{ic}^{SWB}}f_{cool}^{SW}\right]=256 M_{\odot}.
\label{Mcloud}
\ee

Let us estimate the energy budget of IVS. Assuming that the IVS is a boundary 
of the SWB of \vel\ we can estimate its  kinetic energy, assuming an expansion 
velocity $V_{exp}\simeq 13\mbox{ kms}^{-1}$ (\citet{testori06})
\begin {equation}
E_{\rm kin}^{IVS}\simeq \frac{1}{2}M_{IVS}V_{exp}^2\simeq 2\times10^{50}\mbox{ erg.}
\end {equation}

Thermal energy inside of the SWB of \vel\ is limited by conditions of strong Vela SNR shock and the dominance
of the mean density of evaporated clouds in postshock plasma.

Strong shock condition means that the ratio of pressures inside the remnant and bubble and
equivalently the ratio of energy densities should be high $ER^{-3}_{SW}/E_{\rm th}^{SWB}R_{IVS}^{-3}\gg1$ or
\be
E_{\rm th}^{SWB}\ll E\left[\frac{R_{IVS}}{R_{SW}}\right]^3\sim 1\times10^{51}\mbox{ erg}.
\ee
The dominance of evaporated cloud material in the postshock region
$C_{hot}^{SW}=n_{c,hot}^{SW}/n_{ic}^{SWB}\gg1$
means that
$n_{ic}^{SWB}\leq 0.1 n_{hot}^{SW}$ and we take hereafter  $n_{ic}^{SWB}\simeq 10^{-3} \mbox{ cm}^{-3}$ (i.e. $C=10$)
and $T_{ic}^{SWB}\simeq 3\times 10^{6} \mbox{ K}$ as reasonable parameters of the hot (intercloud) gas inside the SWB
enclosed in the IVS. The total mass and thermal energy of this hot gas inside  \vel\  is $M_{hot}^{SWB}\simeq 10 M_{\odot}$
and $E_{hot}^{SWB}\simeq 1\times 10^{49} \mbox{ ergs}$. Meanwhile, the  thermal energy of clumps and clouds with a reasonable value of temperature $T_{cl}\leq 10^4 \mbox{ K}$ is only $E_{cl}^{SWB}\leq 1\times 10^{48} \mbox{ ergs}$.

To summarize, in our hydrodynamical model of IVS the total (thermal and kinetic) energy of the \vel\ SWB is
\be
E_{SWB}=E_{\rm kin}^{IVS}+E_{hot}^{SWB}+E_{cl}^{SWB}\simeq 2\times 10^{50} \mbox{erg}
\label{Etot}
\ee
with evident dominance of kinetic energy of the massive ($\sim 10^5M_{\odot}$) IVS in the total sum.
Inside the SWB we predict about $10 M_{\odot}$ of hot intercloud gas and about $260 M_{\odot}$
of immersed warm/cool clumps/clouds  (in case of their complete evaporation inside Vela SNR). It is worth noting that temperature and density of plasma inside IVS are not well restricted by observations, and here we use approximate values, which are consistent with the adopted limits. 

We can compare these estimates with the results of numerical simulations of stellar wind bubbles around WR stars.
Numerical simulations of the evolution of a star with an initial mass $M_{ini}=35 M_{\odot}$, as proposed by \cite{eldridge09}for WR11, were made by \cite{freyeretal05} for an environment with a density
of $n_0=20\mbox{ cm}^{-3}$ and a temperature of $T_0=200 \mbox{ K}$. They show that at the end of the calculations (before the SN explosion)
the hot gas bubble has a mean radius of $34 \mbox{ pc}$ and shell-like HII and HI regions of the swept up ambient gas extend out to a distance of $43-44 \mbox{ pc}$,
the total mass is  $1.5\times 10^5 M_{\odot}$, the kinetic energy is $4.9\times10^{49} \mbox{ erg}$, the thermal energy of the hot gas
is $1.1\times10^{50} \mbox{ erg}$, of the warm gas  $4.3\times10^{49} \mbox{ erg}$, i.e., the radius, total mass and total energy
(kinetic and thermal, $2\times10^{50} \mbox{ erg}$) are surprisingly close to our estimate for \vel .  Nevertheless, two
important differences should be clarified for \vel\ namely, the kinetic energy dominance and the low density of the ISM.

\subsection{Density of the interstellar medium and interaction with the Gum nebula}

The three estimates for the "typical" ISM density derived above provide widely different values.
The estimate based on the dynamics of expansion of Vela SNR suggests a low value for the ISM density
$n_{\rm ISM}\le 0.01$~cm$^{-3}$, while the estimate based on the total mass of the SWB
around \vel\  ($n_{\rm ISM}\sim 16$~cm$^{-3}$) and the estimate based on the dynamics of expansion of the
SWB ($n_{\rm ISM}\sim 20$~cm$^{-3}$) suggest a much higher density. This points to the fact that
the distribution of the ISM in the direction of the Vela region, in the distance range $\sim 300-400$~pc, is
highly inhomogeneous. This is, in principle, not surprising, because the region is known to contain several
stellar formations with different properties.

First, \vel\ belongs to the $\gamma$Vel association, which is a subcluster of the OB-association Vela OB2
\citep{jeffries08}. The density of the ISM in the OB association soon after \vel's formation is expected
to have been much higher because of the presence of a parent molecular cloud. The initial expansion of the
\vel\ bubble into a dense $\left(10^2-10^3\mbox{ cm}^{-3}\right)$ molecular cloud (the progenitor of
Vela OB2) can explain the large mass of the swept up ISM in a shell around the SWB.
In this scenario, the stellar wind and the radiation of \vel\ destroyed the parent molecular cloud and swept up its gas.
For some time the stellar wind of \vel\ was practically trapped
inside a $\sim 10^5 M_\odot$ cloud, and only when the energy accumulated in the SWB and HII region exceeded
the gravitationally bound energy of the cloud ($\sim 10^{50}\mbox{ ergs}$ for $R_{\rm cl}=10 \mbox{ pc}$ cloud),
the dense shell of swept-up cloud material began to be accelerated by the thermal pressure of the SWB and the HII region gas
without considerable additional mass loading and counter pressure of
hot rarefied gas of the ISM. At this evolutionary stage, the thermal energy of the system converts into the kinetic energy of the shell,
resulting in the atypical dominance of the kinetic energy of the IVS in the total energy balance of the \vel\ SWB/HII region.

\begin{figure}
\includegraphics[width=\linewidth]{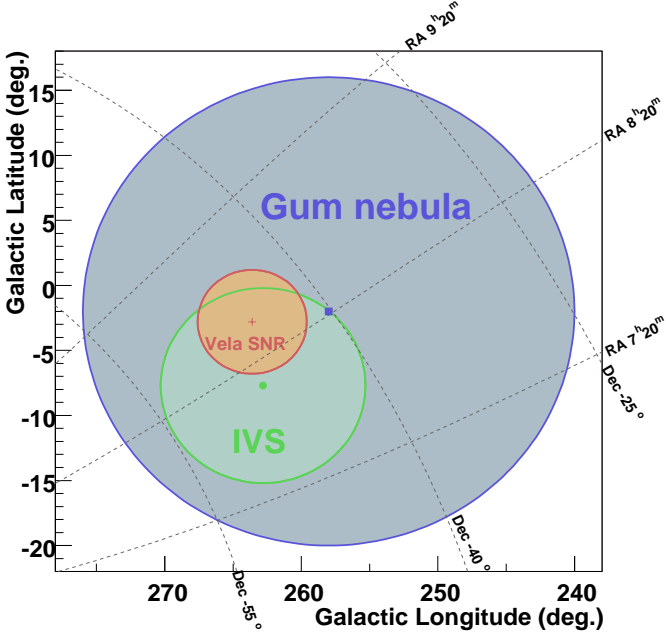}
\caption{Locations of the Vela SNR (Vela pulsar is shown as a cross), \vel\ (shown as a  circle),
IRAS Vela Shell (IVS)  bubble and Gum nebula (center is shown as a square)
in Galactic coordinate system.}
\label{fig:Vela_IVS_Gum}
\end{figure}

Next, both the Vela SNR and the \vel\ SWB could be interacting with a still larger scale SNR,
known as the Gum nebula.  This nebula is a very large region of the ionized gas about
$36^\circ$ in diameter, centered approximately at
$(l,b)=(258^\circ,-2^\circ)$ \citep{gum52}, shown in Fig. \ref{fig:Vela_IVS_Gum}. 
 \citet{brandt71}
explained the Gum nebula as a fossil Stromgren sphere of the Vela X supernova and estimated the distance to nebula  to be
\begin{equation}
D_{\rm Gum}=400\pm 60\mbox{ pc}.
\end{equation}

\citet{reynolds76} suggested that the Gum nebula is an 1 Myr old SNR, which is now heated and ionized by the two very hot stars zeta Puppis and gamma Velorum within it. Later \citet{woermann01} showed that the Gum nebula can possibly be a SNR of the zeta Puppis companion. They showed that runaway O-star zeta Puppis was within $<0.5$ deg of the expansion center of the Gum nebula about 1.5 Myr ago, which is evidence of the relation between the SN explosion of the binary companion and the Gum nebula expansion. Assuming a distance of 400~pc the radius of Gum nebula is about
\begin{equation}
R_{\rm Gum}\simeq 124\left[\frac{D_{\rm Gum}}{400\mbox{ pc}}\right]
\mbox{ pc}.
\end{equation}
This means that both the Vela SNR and \vel\ are situated inside the cavity formed by the expansion of the SNR associated to the Gum nebula.

 The Gum nebula as a very old ($0.9-2.0$~Myr) SNR should be at the late radiative stage of evolution, which can be modeled
with  the solution of \cite{cioffi88} for the shock dynamics. Assuming the the current radius 124 pc,
age $\sim 1.5$~Myr and the explosion energy of $10^{51}E_{51}$ erg, we can find out the density of the
ISM in which the Gum nebula expands:
\begin {equation}
n_{\rm ISM}=0.07\varsigma^{-\frac{5}{36}}\left[ \frac{t}{1.5 \mbox{ Myr}}\right]^{-\frac{7}{6}}
\left[ \frac{D_{\rm Gum}}{400 \mbox{ pc}}\right]^{-\frac{35}{9}}E_{51}^{\frac{17}{72}}\mbox{cm}^{-3},
\end {equation}
where $\zeta$ is the metallicity factor, equal to 1 for solar abundances.

The estimate of the average density of the ISM around Gum nebula shows that actually the nebula expands
into the medium with a much lower density than the one suggested by the estimate of the total mass of the
SWB around \vel. This supports the hypothesis that the enhancement of the density of the
ISM was locally present around the \vel\ system at moment of its birth, most probably because of an OB association.

\section{Conclusions}

We developed a model for the interaction of the Vela SNR and the \vel\ SWB,
which explains the observed NE/SW asymmetry of the Vela SNR.

Adopting a model of the expansion of the Vela SNR into a "cloudy" ISM, we showed that
the volume-averaged density of the shock-evaporated clouds in the NE part of the SNR has
to be about four times higher than in the SW part. We noticed that a plausible explanation
for the observed density contrast is that the Vela SNR exploded  at the boundary of the SWB 
around a nearby Wolf-Rayet star in the \vel\ system, which is situated at approximately
the same distance as the Vela SNR.

Within our model of the interaction of the Vela SNR and the \vel\ SWB, the
difference of the spectral characteristics of the X-ray emission from the NE and SW parts of the
remnant can be used for an estimate of the parameters of the \vel\ bubble. We showed that
the measurement of the change of the column density of the neutral hydrogen gives an estimate of the
total mass of the SWB, $\sim 1\times 10^5M_\odot$.

On the basis of modeling the dynamics of expansion of the bubble around \vel, we confirmed the 
initial mass of the Wolf-Rayet star in the \vel\ system suggested by \cite{eldridge09} to be $\simeq 35M_\odot$.
This estimate is lower than the previous estimates used for the derivations of predictions of the
flux of a $\gamma$-ray spectral line at 1.8~MeV, expected from the decays of $^{26}$Al in this source.
Taking into account the revised estimate of the initial mass of the Wolf-Rayet star,
the $^{26}$Al line flux flux from \vel\ is expected to be much below the COMPTEL limit.

\begin{acknowledgements}
We would like to thank the referee, John Dickel, for many useful comments and suggestions, which appreciably
inproved the paper. IS acknowledges support from Erasmus Mundus, "External Cooperation Window". 
\end{acknowledgements}


\end{document}